\documentstyle[sprocl,epsfig]{article}

\def\beq{\begin{equation}}
\def\eeq#1{\label{#1}\end{equation}}
\def\eeqn{\end{equation}}

\def\beqa{\begin{eqnarray}}
\def\eeqa#1{\label{#1}\end{eqnarray}}
\def\eeqan{\end{eqnarray}}
\def\CR{\nonumber \\ }

\def\leqn#1{(\ref{#1})}

\let\bar=\overbar

\def\etal{{\it et al.}}

\def\ifb{fb$^{-1}$}

\def\L{{\cal L}}

\def\half{\frac{1}{2}}

\def\Dslash{\not{\hbox{\kern-4pt $D$}}}
\def\dslash{\not{\hbox{\kern-2pt $\del$}}}

\def\CM{{\mbox{\scriptsize CM}}}

\def\ee{e^+e^-}
\def\sstw{\sin^2\theta_w}
\def\cstw{\cos^2\theta_w}
\def\mz{m_Z}

\def\mw{m_W}

\def\msb{{\bar{\scriptsize M \kern -1pt S}}}

\def\ELER{e^-_Le^+_R}

\def\s#1{\widetilde{#1}}

\def\Title#1{\begin{center} {\Large #1 } \end{center}}
\def\Author#1{\begin{center}{ \sc #1} \end{center}}
\def\Address#1{\begin{center}{ \it #1} \end{center}}

\def\submit#1{\begin{center} #1 \end{center}}
\def\doeack{\footnote{Work supported by the Department of Energy,
                     contract DE--AC03--76SF00515.}}
\def\SLAC{Stanford Linear Accelerator Center\\
    Stanford University, Stanford, California 94309 USA}
\newcommand\pubblock{\rightline{\begin{tabular}{l} 
         SLAC-PUB-8288\\ October 1999 \end{tabular}}}
\newenvironment{Abstract}{\begin{quotation} \begin{center}
                       ABSTRACT
     \end{center}\bigskip  }{\end{quotation}}

\begin{document}
\begin{titlepage}
\pubblock

\vfill
\Title{Physics Goals of the Linear Collider}
\vfill
\Author{Michael E. Peskin\doeack}
\Address{\SLAC}
\vfill
\begin{Abstract}
I review the most important objectives of the 
physics program of a next-generation $\ee$ linear collider.
\end{Abstract}
\vfill
\submit{Introductory theory lecture\\
presented at the International Workshop on Linear Colliders\\
          Sitges, Barcelona, Spain, 28 April -- 5 May 1999}

\vfill
\end{titlepage}
\def\thefootnote{\fnsymbol{footnote}}
\setcounter{footnote}{0}

\hbox to \hsize{\null}
\newpage
\setcounter{page}{1}

\title{PHYSICS GOALS OF THE LINEAR COLLIDER}

\author{MICHAEL E. PESKIN}

\address{Stanford Linear Accelerator Center\\
    Stanford University, Stanford, California 94309 USA}

\maketitle\abstracts{I review the most important objectives of the 
physics program of a next-generation $\ee$ linear collider.}

\section{Introduction}

For more than twenty years, high-energy physicists have dreamed about using
linear $\ee$ colliders to extend the reach of $\ee$ annihilation to the TeV
energy scale.\cite{Feldman}
About ten years ago, with the first results from the precision
electroweak experimental program at SLC, LEP, and the Tevtron, it became
possible to envision a sharply focused physics program for linear collider
experiments that would begin at center-of-mass energies 
of 400--500 GeV.\cite{mysaariselka,JLC1}
The experimental results of the past few years---in particular, the dramatic 
confirmation of the theory of the electroweak interactions to 
part-per-mil precision---have made the experiments proposed for the
linear collider seem even more urgent and central to the goals of high-energy
physics.

In this article, I will briefly review the most important physics objectives of
the program planned for the next-generation $\ee$ linear collider (LC). 
Recently, a number of detailed reviews have appeared which discuss the broad
array of measurements that can be performed at 
the LC.\cite{purple,MandP,Accom}  My goal 
here is to highlight those measurements that, in my opinion,
form the key justifications for the LC program.

Why do we expect to find new physics at the LC?  The most important 
experimental discovery of the past decade has been the success of the 
Glashow-Weinberg-Salam theory of unified weak and electromagnetic 
interactions.   This model is based 
on the idea that the weak and electromagnetic interactions are mediated by
vector bosons associated with a symmetry group $SU(2)\times U(1)$, which is
spontaneously broken to $U(1)$, the gauge symmetry of Maxwell's equations.
The characteristic  prediction of gauge theory is that coupling constants
should be universal, and, indeed, experiments at the $Z^0$ have shown that 
that the weak and electromagnetic couplings of all
species of quarks and leptons are given by two universal
couplings $g$ and $g'$ (or $e$ and $\sstw$). At the 1\% level of accuracy,
there are deviations from this prediction, but these are accounted for 
by the radiative corrections of the electroweak theory when one uses the
observed mass of the top quark.\cite{EWWG} 

This success brings into relief the fact that
the foundation of the electroweak theory  is shrouded in mystery.  
We have
no direct experimental information on what agent causes the spontaneous
breaking of $SU(2)\times U(1)$ symmetry, and even the indirect indications
are fairly meager.  In the minimal model, this symmetry breaking is due to 
a single Higgs boson, but the true story is probably more complex.
On the other hand, the information must be close at hand.
In the electroweak theory, the formula for the $W$ boson mass is
$\mw = gv/2$,
and from the known value of $g$ we can find the mass scale of electroweak
symmetry breaking: $v = 246$ GeV. Simple arguments from unitarity tell us
that the Higgs boson or some other particle from the symmetry breaking
sector must appear at energies below 1.3 TeV.\cite{LQT}  But, further,
models in which the Higgs boson is very heavy give electroweak radiative
corrections which are inconsistent with the precision experiments.
The analysis of radiative corrections 
requires either that the Higgs boson lie at a mass below 250 GeV,
or that other new particles with masses at about 100 GeV be present to 
cancel the effects of a heavy Higgs boson.\cite{EWWG} 

Unless Nature is very subtle, the first signs of the electroweak symmetry
breaking sector will be found before  the LC begins operation.
  There is a significant
window for the discovery of the Higgs boson at LEP 2 or at  the Tevatron.
In almost every scenario, the Higgs boson or other signals of new physics
will appear at the LHC.  Our problem, though, is not just to obtain some
clues but to solve the mystery.  For this, the unique precision
and clarity of information from the LC will play a crucial role.
 
Because the role of the LC will most likely be to clarify the nature of 
new physics discovered elsewhere, that role depends on what new particles
are observed. In particular, it depends on the actual mechanism 
of electroweak symmetry breaking (EWSB).  To justify the LC project at our
current state of knowledge, one must be prepared to argue that, in any 
model of EWSB, the LC brings important new information that cannot be 
obtained from the LHC.  Systematic  analysis shows that this is the case.
On the other hand, this line of reasoning put a spotlight on
specific precision measurements and requires that the LC
experiments be capable of performing them.  I will point out a number of 
these crucial experiments in this review.

My survey of the LC program will proceed as follows:  First, I will introduce
the capabilities of $\ee$ annihilation experiments by discussing the 
search for contact interactions in $\ee\to f \bar f$.  Next, I will 
review experiments relevant to strong-coupling models of EWSB.  Finally,
I will review experiments relevant to weak-coupling models of EWSB.

\section{Contact Interactions}

Before I discuss detailed models of EWSB, I would like to call attention to 
the ability of the LC to make precise test of the structure of the electroweak
interactions at very short distances.  This study brings in  a number of
unique features that the LC can also use to study more complex reactions
involving new particles.  Here we see these features used in their simplest
context, the study of $\ee\to f \bar f$.

The study of $\ee$ annihilation to fermion pairs begins from the observation
that the Standard Model cross section formulae are simple and depend
only on electroweak quantum numbers.  For example, 
\beqa
&  &\hskip-1.0cm   {d\sigma\over d \cos\theta}(\ELER \to f_L \bar f_R) \ = \ 
 {\pi \alpha^2\over 2 s} N_C \CR & &\cdot
  \biggl| Q_f + {(\half - \sstw)(I^3_f - Q_f \sstw)
   \over \cstw \sstw} {s\over s - \mz^2} \biggr|^2 \cdot (1 + \cos\theta)^2\ .
\eeqa{basicscform}
In this formula, $N_C =1$ for leptons and 3 times the QCD enhancement for 
quarks, $I^3_f$ is the weak isospin of $f_L$, and $Q_f$ is the electric
charge.  The angular distribution is characteristic of annihilation to 
spin-$\half$ fermion pairs.  For $f_L$ production,
the $Z^0$ contribution typically interferes
with the photon constructively for an $e^-_L$ beam and destructively
for an $e^-_R$ beam. Thus, initial-state polarization is a useful diagnostic.
 For annihilation to the $\tau$ and  the top quark,
the final state polarization can also be measured. 

\begin{figure}
\begin{center}
\leavevmode
{\epsfxsize=4.50truein \epsfbox{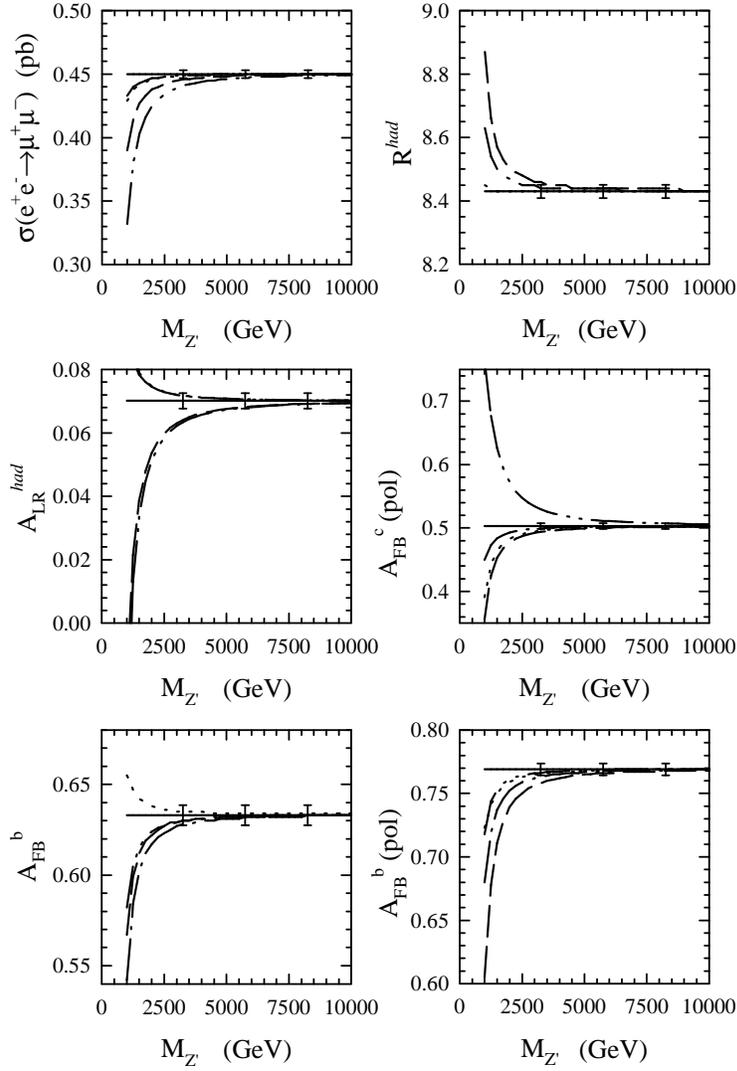}}
\end{center}
 \caption{Deviations from Standard Model predictions for various
    $\ee\to f\bar f$ processes due to a new $Z^0$ boson, from a 
       study by Godfrey, ref. 9.
  The error bars shown correspond to a 50 \ifb\ event sample at 
   $E_\CM = 500$ GeV.}
\label{fig:Godfrey}
\end{figure}

This simplicity of formulae such as \leqn{basicscform} allow one to determine
unambiguously the spin and Standard Model  quantum numbers of any new
state that is pair-produced in $\ee$ annihilation.  Applied to the familiar
particles, they provide a diagnostic of the electroweak exchanges that might
reveal new heavy weak bosons or other types of new interactions. These 
tests can be applied independently to the couplings to $e$, $\mu$, polarized
$\tau$, $c$, $b$, and light quarks.  Figure \ref{fig:Godfrey}
 illustrates
how the available set of observables can be used to study the couplings of
a new $Z^0$ in four different models for its couplings.\cite{Godfrey}  

A  1 TeV linear collider would be sensitive, through these precision 
measurements, to a new $Z^0$ up to masses of about  4 TeV. 
 A new $Z^0$ boson would also appear
at the LHC, up to a similar reach in mass,  as a resonance in $e$ or $\mu$
pair production.  However, little can be learned about its couplings
if its mass is above about 1 TeV.  For such a boson, the LC will fill in 
the picture of its couplings to quarks and leptons.  Measurements of 
simple annihilation processes can also be 
used to test for new interactions that would signal quark and
lepton compositeness; a 50 \ifb\ event sample at 500 GeV would be sensitive
to a compositeness scale  $\Lambda$ of 30 TeV.\cite{HGP}  More exotic
effects are also possible.  Recently proposed
 models with large extra dimensions predict
contact interactions due to graviton exchange.  These precision measurements
can not only reveal the presence of these interactions, but also their
spin-2 character.\cite{Hewett}

\section{Strong-Coupling Route to EWSB}

In the remainder of this article, I will focus on topics relevant to 
the question of electroweak symmetry breaking (EWSB).  As I have explained
above, the origin of electroweak symmetry breaking must lie in the TeV energy
region.  In principle, EWSB could either be generated by a weak-coupling 
theory with an elementary Higgs boson or by a strong-coupling theory, with
the symmetry-breaking possibly due to a composite operator.  Many models 
have been proposed that illustrate the two  viewpoints.  The models of the 
two classes have quite different phenomenological implications.

I will first consider models of EWSB due with  strong-coupling dynamics.
In such models, the signals of the EWSB mechanism are most clear in the
properties of the heaviest Standard Model particles, the $W$ and $Z$ bosons
and the top quark.  The LC can illuminate this mechanism through its ability
to study the couplings of these particles in detail.  Often, the 
model of EWSB will also contain new particles that decay to weak bosons and
third-generation fermions.  The LC would allow these particles to be 
studied by the same techniques.

\subsection{$W$ boson}

Consider first the $W$ boson.  The process $\ee\to W^+W^-$ is the most 
important 
single process contributing to 
 $\ee$ annihilation at high energy.  This process also
has numerous features that make it especially amenable to detailed study.

From the viewpoint of EWSB, the $W$ is interesting because it receives mass
through the Higgs mechanism.  The massless $W$ has only two degrees of freedom,
corresponding to transverse polarizations.  The massive $W$ has a third
degree of freedom, which corresponds to 
the longitudinal polarization state.  This state must be
stolen from the symmetry-breaking sector.  In fact, it is a theorem 
in quantum field theory that, in the limit of high energy, the amplitude
for producing a longitudinally polarized $W$ is given precisely by the 
amplitude for producing the charged
Goldstone boson associated with $SU(2)\times U(1)$ 
symmetry-breaking.\cite{GBET}

Effects of new physics on the cross section for $\ee\to W^+W^-$ are
traditionally expressed in terms of effective 3-vector boson couplings
$g_{1Z}$, $\kappa_{\gamma,Z}$, $\lambda_{\gamma,Z}$.  These in turn are
given in terms of  coefficients $L_i$ that appear in the effective Lagrangian
describing the Goldstone bosons.\cite{Strong}  The parameter 
deviations predicted are rather small; for example, new strong interactions
similar to QCD at TeV energies would give a deviation $(\kappa_\gamma-1)
\sim 3\times 10^{-3}$.  This should be compared with upper limits 
of several percent which have been obtained from LEP 2.\cite{LEP2TGV}

To do better, the LC can take advantage of several features.  First, the
effect of the Goldstone boson couplings is naturally 
enhanced by a factor $s/\mw^2$.
Second, going to higher energy separates the $W^+$ and $W^-$ into opposite
hemispheres and makes the kinematics more well-defined.  In 
Figure~\ref{fig:Miyamoto}, I show the results of a simulation 
study of events at a 500 GeV LC 
in which one $W$ decays hadronically and the other 
leptonically.\cite{Miya}  The full detail of the 
reaction, including both production and decay angles, can be reconstructed.
In particular, the $W$ bosons at  central values of the decay angle
 $\cos\theta$ are those
with longitudinal polarization.  By fitting the full multi-variable
distribution, it is 
possible to obtain limits on the $\kappa$ and $\lambda$ parameters at the 
$10^{-3}$ level at 500~GeV, and even more stringent limits at higher
energy.\cite{Aihara}

\begin{figure}
\begin{center}
\leavevmode
{\epsfxsize=3.40truein \epsfbox{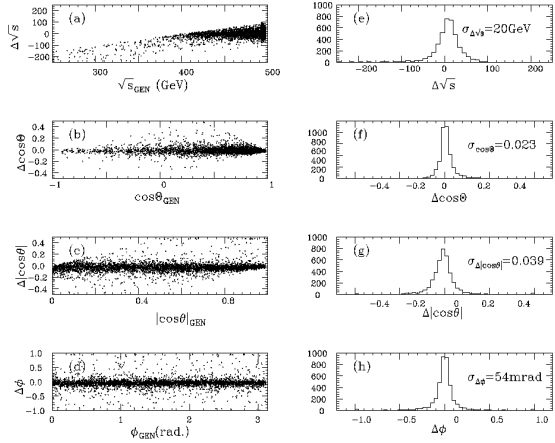}}
\end{center}
 \caption{Reconstruction of production and decay angles in $\ee\to W^+W^-$,
 from a simulation study by Miyamoto, ref. 15}
\label{fig:Miyamoto}
\end{figure}

\subsection{$WW$ scattering}

The principle that gives us access to the production amplitudes for 
states from the symmetry-breaking sector also allows us to study the 
interactions of these particles.  In the reactions $\ee \to \nu \bar \nu
VV$, where $VV$ is $W^+W^-$ or $Z^0Z^0$, the most important subprocess is 
that in which the incoming electron and positron radiate a $W^-$ and $W^+$,
which then collide and scatter.  One can show that a substantial
fraction of the  radiated $W$'s are longitudinally 
polarized.\cite{CD}  The 
scattering amplitudes for these bosons come directly  from the 
symmetry-breaking interactions.

Experiments on these scattering processes
 are difficult both at the LC and at the LHC.  At the LHC, one can 
radiate $W$'s from quark lines, detect the final vector bosons using their
leptonic decays, and apply a forward jet tag or other topological cuts to 
enhance the signal over background.
At the LC, one can study vector bosons using  their
hadronic decay models, imposing a cut on the total transverse momentum of
the $VV$ system to remove background from two-photon processes.  
It is important to be able  to separate $W$ and $Z$
on the basis of the 2-jet mass.\cite{Han}
Table~\ref{tab:StrongTable}, taken from ref. 13, compares the 
capabilities of LHC and the LC for 100 \ifb\  event samples and an assumed
LC energy of 1.5 TeV.  (A larger LC luminosity sample would
allow the study to be done at somewhat lower energies.) 
A notable advantages of the LC is its extraordinary
sensitivity to vector resonances, which show up as $s$-channel resonances 
in  $\ee\to W^+W^-$.  The LC also has a unique advantage in its ability 
to study the reaction $W^+W^- \to t\bar t$,\cite{Barklowtt} a reaction 
that directly probes the coupling of the top quark to the symmetry-breaking 
sector.

\begin{table}[t]
\begin{center}
\caption{Estimated LHC and NLC sensitivity to resonances in the new strong
interactions. For details, see ref. 13.} 
\label{tab:StrongTable}
\begin{tabular}{cccccc}
\hline
\hline
Machine & Parton Level Process & I & Reach & Sample & Eff. $\L$ Reach \\ 
\hline \\
LHC & $qq' \to qq'ZZ$ & 0 & 1600 & $1500^{+100}_{-70}$& 1500 \\ \\
LHC & $q \bar q \to WZ$ & 1 & 1600 & $1550^{+50}_{-50}$ & \\ \\
LHC & $qq' \to qq'W^+W^+$ & 2 & 1950 & $2000^{+250}_{-200}$&  \\ \\
NLC & $e^+e^- \to \nu \bar \nu ZZ$ & 0 & 1800 & $1600^{+180}_{-120}$&
 2000 \\ \\
NLC & $e^+e^- \to \nu \bar \nu t \bar t$ & 0 & 1600 & $1500^{+450}_{-160}$&
\\ \\
NLC & $e^+e^- \to W^+W^-$ & 1 & 4000 & $3000^{+180}_{-150}$ \\ \\
\hline
\hline
\end{tabular}
\end{center}
\end{table}

\subsection{Top quark}

Finally, the LC can access a strongly-coupled symmetry breaking sector 
through precision studies of the heaviest Standard Model particle, the top
quark.  The pair production reaction $\ee\to t \bar t$ may be studied either
at threshold or at higher energy.

The Standard Model prediction for  $\ee\to t \bar t$, like the prediction 
for  $\ee\to W^+W^-$, has a rich structure.  The production cross section
depends strongly on both the electron and the t quark polarization. For
example, the subprocess $\ELER \to t \bar t$ is dominated by forward 
production of $t_L$.  The top polarization is visible because the 
short $t$ lifetime guarantees that a produced $t$ will not be depolarized
by soft hadronic interactions,\cite{Rapallo} and because the dominant
decay $t\to b W^+$ and the subsequent $W^+$ decay have distributions 
sensitive to polarization.  To take advantage of the final-state polarization
observables, it is necessary to be able to reconstruct $t\bar t$ events
efficiently in the 6-jet mode produced by hadronic $W$ decays on both 
sides.\cite{Fujiitop}

In a theory with strong-coupling electroweak symmetry breaking, the top
coupling to the strong sector shows up in its coupling to gauge bosons.
Already in the Standard Model,  70\% of the $W^+$'s from top decay are
longitudinally polarized, reflecting the dominance of the top Yukawa coupling
over the $SU(2)$ gauge coupling in top decays.  This fraction may be enhanced
in strong-coupling models.  In technicolor models, the $Z^0$ coupling to 
third-generation quarks is predicted
 to be shifted by diagrams involving extended
technicolor boson exchange.  This effect is not seen in the $Z^0\to b \bar b$
coupling.  However, it is natural that effects which cancel in that coupling
add constructively in the coupling to top, giving rise to shifts of
up to 10\% in  the
$Z^0 t \bar t $ coupling that would be revealed by the measurements of the 
polarization asymmetry for top production.\cite{ZZtop}  On the other hand,
if there is a light Higgs boson, it should be possible to observe the 
process $\ee\to t\bar t h^0$ and thus measure the $t\bar t h$ coupling 
directly.

It is also interesting to obtain as accurate as possible a value for the 
top quark mass, both because of the important role of virtual top quarks in 
phenomonology and because of its intrinsic interest for the problem of
flavor.  At a LC, the top quark mass can be computed from the position of
the $t\bar t$ threshold.  The energy region that, for lighter quarks, holds
the bound quarkonium states is smeared out by the large top quark width.  
The resulting smeared shape can be computed accurately in QCD.  The position
of the threshold can be located to about 200 MeV with relatively small
data samples (10 \ifb), given an accurate value of $\alpha_s$.  The
results of a simulation study are
 shown in Figure~\ref{fig:sumino}.  The threshold position can be
related to the  short distance parameter $m_{t \msb}(m_t)$ with 
a similarly small error.\cite{Hoang}

\begin{figure}
\begin{center}
\leavevmode
{\epsfxsize=4.50truein \epsfbox{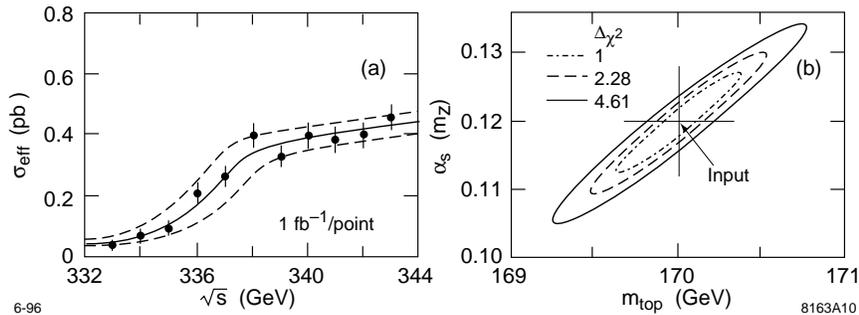}}
\end{center}
 \caption{Measurement of the  $t \bar t$ threshold location, from a 
      simulation study by Sumino, ref. 23.}
\label{fig:sumino}
\end{figure}

\section{Weak-Coupling Route to EWSB}

The alternative class of models of electroweak symmetry breaking are those
in which $SU(2)\times U(1)$ is broken by the vacuum expectation value of a 
weakly-coupled Higgs scalar field. In these models, there is a light Higgs
boson, and possibly also a spectrum of heavier Higgs states. 
Since the precision electroweak data favor a low
Higgs boson mass and also exclude large modifications of the $Zb\bar b$
coupling, it is this alternative which currently
 has the most experimental support.  A Higgs boson in this mass range 
should be discovered before the LC experiments, at LEP 2 or the Tevatron
and certainly at the LHC.  However, it will be the LC that tests whether
this particle indeed  generates the  quark, lepton, and gauge boson masses.

The simplest weak-coupling models do not explain why $SU(2)\times U(1)$ is 
broken.  Rather, the symmetry-breaking is the result of a negative (mass)$^2$
parameter for the Higgs field that is inserted into the Lagrangian by hand.
The only way to avoid this unsatisfactory situation without requiring 
strong coupling is to introduce a 
symmetry that links the Higgs field to some field of higher spin. This 
eventually requires that the theory of electroweak symmetry breaking be
supersymmetric.  Conversely, a supersymmetric generalization of the Standard
Model easily generates a symmetry-breaking potential for the Higgs field as
the result of radiative corrections due to the heavy top quark.  Thus, 
the assumption that EWSB has a weak-coupling origin leads naturally to 
supersymmetry.

Both aspects of the weak-coupling models have interesting implications for 
the LC.  The light and heavy states of the Higgs boson spectrum can 
be studied in detail in $\ee$ annihilation. The LC also offers many 
incisive tools for the precision study of the spectrum of supersymmetric 
particles.

\subsection{Higgs boson}

One of the key aspects of the LC experimental program is the study of a 
light Higgs boson.  Any Higgs boson with a mass below 350 GeV can be 
studied at a 500 GeV LC through the reaction $\ee\to Z^0 h^0$.  Though the 
Higgs boson is not produced at rest as a resonance, the experimental setting
is extremely clean.  The $h^0$ appears as a peak at a definite recoil energy,
and our precise knowledge of the $Z^0$ mass and branching ratios can be used
to establish the signal in a variety of $h^0$ decay modes.

The crucial question for a light Higgs boson is, does it couple to all 
species proportional to mass?  To test this, one may check the relative
Higgs branching ratios predicted by the Minimal Standard Model.
The relative rates for $b$, $c$, and $\tau$ pairs 
(72\%:3\%:7\% for $m_h = 120$ GeV) correspond to an 
identical scale for the Higgs couplings
 to down quarks, up quarks, and leptons.  
In multi-Higgs models, the lightest Higgs will typically couple preferentially
either to up- or to down-type fermions.
The coupling to $WW$ and the total $Zh$ production
rate, which is proportional to the $hZZ$ coupling, test the extent to which
the
$W$ and $Z$ masses that are due to the $h^0$.  The branching
ratios to $gg$ and $\gamma\gamma$ measure sum rules over the colored and 
uncolored massive spectrum.\cite{GunHaber} In Figure~\ref{fig:Higgs},
I show a recent estimate of the accuracies that can be achieved in a 
variety of Higgs decay modes.\cite{Battaglia}  

\begin{figure}
\begin{center}
\leavevmode
{\epsfxsize=2.80truein \epsfbox{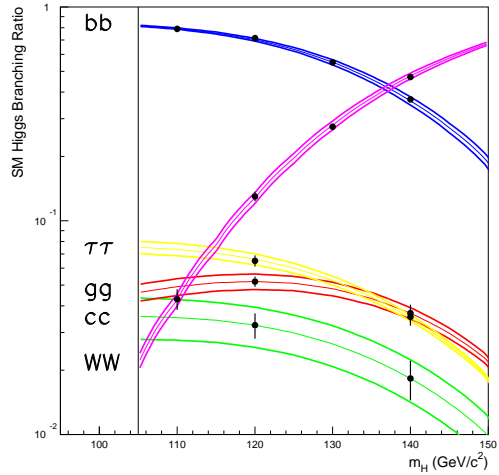}}
\end{center}
 \caption{Measurement of the Minimal Standard Model Higgs boson 
 branching ratios, from a simulation study by Battaglia, ref. 25,
 assuming 500 \ifb\ at $E_\CM = 350$ GeV.}
\label{fig:Higgs}
\end{figure}

The measurement of the 
$\gamma\gamma$ branching ratio or partial width from $Zh$ production
requires very large luminosity samples.  Alternatively, this measurement 
is straightforward at a $\gamma\gamma$ collider and provides a strong
physics motivation for developing that technology.\cite{Borden,Hgg}

There is no reason why a weakly-coupled Higgs sector should not contain
several scalar fields whose vacuum expectation values contribute to the
$Z$ and $W$ masses.  Experiments at the LC can discover the 
complete set of these
bosons and prove that they are fully responsible for the vector boson
masses.  To be specific, let the vacuum expectation value of the $i$th Higgs 
$h_i^0$ be $f_i v$, where $v = 246$ GeV.   Then the $h^0_i$ is  produced
in recoil against the $Z^0$ with a cross section equal to a factor 
$f_i^2$ times the cross section for a Minimal Standard Model Higgs of that
mass.  We have found the
 full set of scalars when the observed bosons saturate the
sum rule~\cite{GHsumrule}
\beq
              \sum_i f_i^2 = 1 \ .
\eeq{GHsumrule}
The ability of the LC to recognize the Higgs boson as a peak in the
$Z^0$ recoil energy spectrum, independently of the Higgs decay mode, is 
crucial for this study.

Models with additional Higgs fields also contain additional heavy spin-0
states. Supersymmetric models, for example, typically 
contain heavy Higgs states that are pair-produced via $\ee\to H^0 A^0$,
$\ee\to H^+H^-$.  The couplings  of these states to fermion pairs are not
universal among species but rather depend strongly on the underlying parameters
of the Higgs sector.  Thus, the branching ratios can be used systematically
to determine these parameters, such as $\tan\beta$, which are needed as 
input in other aspects of the theory.\cite{FengMoroi}

\subsection{Supersymmetry}

I have explained above that supersymmetry is naturally connected to the idea
of weak-coupling electroweak symmetry breaking.  Many theorists (I am one)
would claim
that any plausible model with a light Higgs boson must contain 
supersymmetry at the TeV scale. 

If supersymmetry is responsible for electroweak symmetry breaking, 
supersymmetric particles should be discovered at LEP2, the Tevatron, or 
the LHC before the LC experiments begin.  Very clever
methods have been devised to make precise mass measurements of supersymmetric
particles at the LHC.\cite{HandPaige} But nevertheless, there are 
intrinsic difficulties in studying supersymmetry at hadron colliders.
It is not possible to determine the initial parton energies or, because of 
unobserved final particles, to reconstruct the complete final state.
All possible supersymmetric particles are produced at once in the same
event sample, so that individual particles must be separated on the basis
of branching to characteristic decay modes.

The LC brings new tools that can clarify the nature of these new particles.
First of all, since cross sections in $\ee$ annihilation depend in a 
model-independent way on the spins and $SU(2)\times U(1)$ quantum numbers
of the produced particles, the LC can verify that new particles have the 
correct quantum numbers to be supersymmetric partners of Standard Model states.
By adjustment of the center-of-mass energy and polarization, one can select
specific states preferentially.  An example is given in 
Figure~\ref{fig:goodman}, where the masses of 
the distinct supersymmetric partners of $e^-_L$ and $e^-_R$ are 
determined by the positions of kinematic endpoints observed 
in $\ee\to \s e^+\s e^-$ with a polarized $e^-$ beams.\cite{goodman}
A detailed analysis in which this strategy is used to make a precise
spectrum measurement is presented in ref. 33.
In systems where superpartners naturally mix---for example, the 
$\s t_L,\s t_R$ and $\s w^+,\s h^+$
combinations---the dependence 
on beam polarization can be used to  measure the mixing 
angles.  For the $\s\tau$ and other states that decay to
$\tau$, the kinematic contraints allow final-state $\tau$ polarization
to be used also as a powerful probe.\cite{fujiinojiri}

\begin{figure}
\begin{center}
\leavevmode
{\epsfxsize=2.80truein \epsfbox{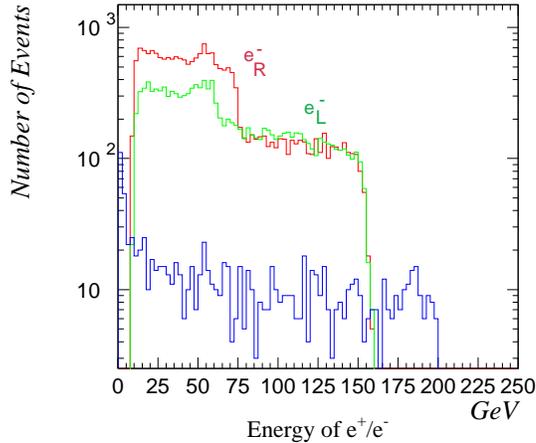}}
\end{center}
 \caption{The electron and positron  energy spectrum in $\ee\to \s e^+ \s e^-$,
 for beams with 80\% $e^-_L$ and $e^-_R$ polarization, from a simulation 
study by Danielson and Goodman, ref. 32.  The lower histogram represents
the background from 2-photon processes.}
\label{fig:goodman}
\end{figure}

These probes are needed because supersymmetry models are typically complex,
with not only a doubling of the particle spectrum but also a number of 
new phenomena.  As one example, I have already noted that, in models of 
supersymmetry, EWSB may arise as
 a byproduct of the renormalization of the scalar
mass spectrum.  We need to be able to measure the underlying parameters of
responsible for this effect to see whether this in fact is the explanation
for EWSB.  In the simplest models, the masses derived from supersymmetry
breaking are independent of flavor, but this is not necessary and must
be tested directly.  In Table \ref{tab:SUSYTable}, I have made a 
more complete list of issues that must be probed experimentally before we
can claim that we understand the supersymmetric generalization of the Standard
Model.  Underlying all of these issues is the question of the origin of 
supersymmetry breaking.  This phenomenon, which supplies most of the new
parameters of a supersymmetric model, would probably
arise from energy scales far above 1 TeV.  The understanding of the new
parameters of supersymmetry could then potentially give us a window into
physics at extremely short distances.\cite{myKyoto}

\begin{table}[t]
\begin{center}
\caption{Questions for the experimental program on supersymmetry}
\label{tab:SUSYTable}
\medskip
\begin{tabbing}
 AAAAAAAA \= AAA \= AAA \=   \kill
 \>$\bullet$ Is it really SUSY?     \\
\>\>$\circ$ new particle quantum numbers, spin, statistics\\
 \>\>$\circ$ identification of complete $SU(2)\times U(1)$ multiplets\\
\>\> $\circ$ SUSY relation of coupling constants\\
\>$\bullet$  Major spectrum parameters\\
\>\>$\circ$   gaugino/Higgsino mixing\\
\>\>$\circ$   gaugino mass ratios: $m_1:m_2:m_3$\\
\>\>$\circ$  flavor universality of $\s q$, $\s \ell_R$, $\s \ell_L$ masses ?\\
\>\>$\circ$    $\s q:\s \ell_R:\s \ell_L$ mass ratios\\
\>\>$\circ$   signatures of gauge- or anomaly-mediation\\
\>\>$\circ$  signatures of R-parity violation\\
\>$\bullet$ Third generation and EWSB\\
 \>\>$\circ$  determination of $\mu$, $\tan\beta$\\
 \>\>$\circ$  mixing of $L/R$ partners for $\s t$, $\s b$, $\s \tau$\\
 \>\>$\circ$  $h^0$ mass\\
 \>\>$\circ$  $H^0$, $A^0$, $H^+$ masses and branching ratios\\
\>$\bullet$ Precision effects\\
\>\>$\circ$   $\s q_L - \s q_R$, $\s u_R - \s d_R$ mass differences\\
 \>\>$\circ$  radiation corrections to coupling relations\\
\>\>$\circ$   slepton flavor mixing\\
 \>\>$\circ$  phases in soft parameters, CP violation
\end{tabbing}
\end{center}
\end{table}

\subsection{Extra dimensions}

Many  people express the opinion that supersymmetry, with large number
of postulated new particles, is too daring a generalization to be the 
true theory of the TeV scale physics.  My own opinion is that it is not
daring enough.  Supersymmetric models require all of their complex components
to explain the details in Nature which are missing from the Standard Model.
But these components are not unified by a common underlying idea.
Contrast with it the theory which is now understood for the GeV scale.  Here
experiment revealed a complex array of new states and couplings, but these 
turned out all to arise from the underlying simplicity of the Yang-Mills
gauge interaction.

Recently, there has been much discussion of a grander idea for the nature
of TeV-scale physics.  For many years, string theory has suggested that 
space-time has more than four dimensions.  It is possible that the scale
of these dimensions, or even the scale of quantum gravity, is as low as 
TeV energies.\cite{Anton,Lykken,ADD}  In this picture, high-energy 
experiments would reveal not only supersymmetry but also the 
higher-dimensional spectrum with extended supersymmetry that characterize
string theory at short distances.

As one might expect, theories with new space dimensions suggest new
phenomena that could be discovered at high energy.\cite{Hewett,GWR}
A low quantum gravity scale would allow gravitational radiation to be seen
in high-energy collisions, both as missing-energy processes and as 
spin-2 contact interactions in fermion-fermion scattering.  A TeV scale
for new dimensions would imply recurrences of the Standard Model gauge
bosons, which would appear as dramatic $s$-channel resonances.  Some of these
phenomena could be observed at the LHC, but many of the new effects would
require the probes with beam polarization and precision measurement which 
are the domain of the LC.

\section{Conclusions}

The success of the Standard Model in accounting for the detailed properties
of the strong, weak, and electromagnetic interactions leads us to focus
attention on physics of electroweak symmetry breaking.  At this time
we do not know the what new physics is responsible for this symmetry breaking.
But, in any scenario, physicists would look to the LC for tools essential
to understanding the new phenomena.  These include the ability to predict 
background cross sections precisely, to interpret signal cross sections
unambiguously, to detect $b$, $c$, and $\tau$ with high efficiency, and to 
analyze the effects of polarization both in the initial state and in 
decays.  The capabilities of the LC will  allow us to characterize
these new interactions in detail, and to uncover their origin.

\section*{Acknowledgments}
This article attempts to summarize a huge body of work and many insights
that have come out of the international study of linear collider physics.
I would like to give special thanks to Charlie Baltay, Sachio Komamiya, and
David Miller for their harmonious organization of the current phase of this
study, and to Enrique Fernandez and the local organizers of the Sitges
meeting for providing such a pleasant setting for this meeting.  This 
work was supported by the US Department of Energy under contract
DE--AC03--76SF00515.

\end{document}
